\documentclass[prl,twocolumn, showpacs,floatfix,amsfonts]{revtex4}

\usepackage{graphicx,graphics,color,epsfig}
\usepackage{bm}
\usepackage{amsmath}
\usepackage{amssymb}

\newcommand{\bk}{{\bf k}}

\newcommand{\br}{{\bf r}}

\newcommand{\bsig}{\mbox{\boldmath{$\sigma$}}}

\begin{document}
\preprint{}
\title{Inelastic Tunneling Spectroscopy  in a D-wave Superconductor.}

\author{ A.V. Balatsky, Ar. Abanov, and Jian-Xin Zhu}
\affiliation{Theoretical Division, Los Alamos National Laboratory,
Los Alamos, New Mexico 87545}
\date{\today}

\begin{abstract}
{ We propose a mechanism to use inelastic tunneling spectrosopy
STM
 to detect  a single spin   in
  a d-wave superconductor and in a pseudogap state,
   based on a direct exchange coupling $J$  between the surface
electrons and the local  spin $\bf S $ in a magnetic field. This
coupling will produce a kink in a $dI/dV$ characteristic at Zeeman
energy of the spin $\omega_0$. We find that for relevant values of
parameters signal scales as $dI^2/dV^2 \simeq (JN_0)^2 \Theta(eV -
\omega_0)$ and could be in the range of $10^{-2}$ of the bare
density of states where $N_0$ is the density of states for surface
electrons. Scattering in  superconductor with the  coherence peak
at gap maximum $\Delta$ leads also to strong features at $ \Delta
+ \omega_0$. This suggests a new technique for a detection of a
local spin excitation with STM. We also consider a detection of a
local vibrational mode as a simple extension of the spin case.}
\end{abstract}

\maketitle

Inelastic electron tunneling   STM spectroscopy   (IETS-STM) is a
well established technique that  has been proven, starting with
important experiments of Stipe et al.~\cite{Ho}. In these
experiments a step-like feature in tunneling current and local
density of states have been observed. The physical explanation of
the effect is straightforward: once energy of tunneling electrons
exceeds the energy required to excite local vibrational  mode,
there is a new scattering process that contributes to the
scattering  of electrons due to inelastic excitation of the local
mode \cite{Scalapino}. Similarly, in case of a single impurity
spin $S$ in external field the localized spin state will be split
with the Zeeman gap $\hbar \omega_0  = g \mu_{B} B$. If there is a
local exchange coupling that allows conduction electrons to
scatter inelastically off the local spin in an external magnetic
field then one has  a very similar situation of inelastic
scattering off localized spin levels. Up to date  the single spin
inelastic tunneling spectroscopy in metals with STM has not been
observed however. One reason that is often mentioned is a Kondo
screening of a magnetic spin by conduction electrons that leads to
a more complicated response~\cite{Appelbaum}.

Potential applications of the
  techniques that are
sensitive to the single spin dynamics include studies of a single
spin Kondo problem~\cite{Manoharan},
 studies of magnetic impurity states in
 superconductors~\cite{Yazdani} and single spin detection and manipulation in a
context
  of a solid state quantum computing schemes~\cite{Kane,q-dots}.
Use of STM for spin detection is a promising  approach that would
allow one to combine a sub-Angstrom spatial
 resolution with the high electronic sensitivity.

 We
propose to study the inelastic electron tunneling spectroscopy of
a {\em single spin} and of a {\em localized vibrational mode} with
STM in a d-wave superconductor. In case of spin  there is a
crucial difference between metal and d-wave superconductor that
might make the single spin observation in IETS with STM more
feasible: vanishing DOS in d-wave superconductor drastically
suppresses or even makes the Kondo temperature $T_K = 0$ for a
single impurity spin. This leaves one  with a  simpler problem of
a scattering off the single unscreened spin in a media with
linearly vanishing density of states (DOS) $N(\omega) \sim
\omega$.

The proposed  approach to identify magnetic sites  is based on the
fact that the local density of states in the vicinity of a single
impurity spin will have a kink-like feature with threshold at $eV
= \hbar \omega_0$, $\omega_0$ is a  Larmor frequency of a spin.
This would be a natural, albeit not tried yet in correlated
electron systems, extension of the single molecule vibrational
spectroscopy and could allow a single spin detection. Our results
can be summarized as follows: i) we find that   spin produces the
kink-like singularity in the density of states that as a function
of position with respect to moment site at low $T \ll \omega_0,
\omega \ll \Delta$ is :
\begin{eqnarray}\label{EQ:1}
 & \delta N(\br,\omega)/N_0 \simeq
2 \pi^2 (\frac{\omega}{\Delta} \ln(\frac {4 \Delta}{\omega}))^2
\nonumber
\\  & \times \frac{\omega - \omega_0}{\Delta} \Theta(\omega -
\omega_0)(N_0JS)^2 \Lambda({\bf r})\;,
\end{eqnarray}
where $\Lambda({\bf r}) $ is a combination of Greens function of
an electron in real space (assumed 2D), describing  Friedel
oscillation as a function of $r$ and reflecting four-fold
anisotropy due to superconducting gap. The singularity at $\omega
= \omega_0(B)$ changes as a function of applied external magnetic
field since $\omega_0 = g \mu_B B$, $\mu_B$ being the Bohr
magneton and $g$ the gyromagnetic ratio; We also find a strong
feature at energies $ \Delta + \omega_0$ as a result of sharp
coherence peak in DOS of a superconductor, $\omega \simeq
 - \Delta$:
\begin{eqnarray}\label{EQ:2}
\delta N(\br,\omega)/N_0 \simeq  \ln^2(\frac{4\Delta}{|\omega -
\Delta|})\ln(\frac{4\Delta}{\omega - \omega_0 + \Delta})
\nonumber\\ \times \Theta(\omega - \omega_0)(N_0JS)^2 \Lambda({\bf
r})\;.
\end{eqnarray}
(see Eq.~(\ref{EQ:DOS2}));  these kink-like singularities in $
dI/dV$ characteristic of a STM tunneling in a vicinity of magnetic
site lead to a  {\em step} in $d^2 I/dV^2$; ii) the strength of
the effect is of second order in a dimensionless coupling $N_0 J$.
If we take typical values of $J \sim 1-0.1 eV$ and $N_0 = 1/eV$ we
find that the magnitude of the correction to DOS is on the order
of $10^{-2}$. iii)The proposed effect can also be trivially
expanded to be tried for a molecular spectroscopy of a local
vibrational mode. One would have to assume $\omega_0$ be the
eigenfrequency of a local mode, independent of the field, and
replace $JS$ by the coupling constant to the local vibrational
mode in Eq.~({\ref{EQ:1}).

Important potential application of the proposed technique  is the
study of the induced magnetic moment near $Zn$ and $Ni$ impurities
in the high-T$_c$ materials. It has been argued that the $Zn$,
$Ni$ and $Li$ impurities in the Cu-O plane generate uncompensated
spin. A $Zn^{2+}$ that substitutes $Cu^{2+}$ has a closed shell
and is nonmagnetic and induced moment near $Zn$  has to be a
collective response  of the neighbor sites. Claims also have been
made about evidence of the Kondo effect~\cite{Alloul}. Direct and
independent test of magnetism induced by $Zn$, $Ni$, and $Li$
impurities would be important for our understanding of the physics
of strong correlations in Cu-O planes in high-T$_c$ compounds. The
IETS STM technique would allow the direct and alternative approach
to distinguish between magnetic and nonmagnetic sites in a d-wave
superconductor.

Assume that we have localized magnetic atom with spin $S$ on a
surface of a d-wave superconductor. Electrons in a superconductor
interact with the localized spin via point-like exchange
interaction at one site $J {\bf S} \cdot \bsig$:
\begin{eqnarray}\label{eq:Ham}
H = \sum_{\bk} c^{\dag}_{\bk \sigma} \epsilon({\bk}) c_{\bk
\sigma} + \sum_{\bk} (\Delta(\bk) c^{\dag}_{\bk
\uparrow}c^{\dag}_{-\bk \downarrow} + h.c.) \nonumber
\\+ \sum_{\bk,\bk', \sigma, \sigma'} J {\bf S}\cdot c^{\dag}_{\bk \sigma}
\bsig_{\sigma \sigma'} c_{\bk' \sigma'} + g \mu_B {\bf S}\cdot
{\bf B}\;,
\end{eqnarray}
where $c_{\bk \sigma}$ is annihilation operator for the conduction
electron of spin $\sigma$, $\epsilon(\bk)$ is the energy of the
electrons, $\Delta(\bk) = \Delta (\cos k_x  - \cos k_y)$ is the
d-wave superconducting gap of magnitude $\Delta \simeq 30 meV$ in
typical high-T$_c$ materials.  The local spin ${\bf S}$ is a
$|S|=1/2$. We focus here on the effect of the Zeeman splitting of
the otherwise degenerate local spin state in the external magnetic
field $B$ with splitting energy $\omega_0 \equiv \omega_L = g
\mu_B B$. Below we use a mean field description of superconducting
state at low temperatures $T\ll T_c$. Assuming field $B \ll
H_{c2}$ we will ignore the orbital and Zeeman effect of the field
on the conduction electrons~\cite{Comment1}.

  We
are  interested in a local effect of  inelastic scattering of
electrons. Thus  only local properties will determine the
conduction electron self-energy. Results we obtain will also hold
for a normal state with linearly vanishing DOS, such as a
pseudogap state of high-Tc superconductors.  In the  case of a
normal state one would model normal pseudogap state  with a single
particle Hamiltoninan $ H_0 = \sum_{\bk} c^{\dag}_{\bk \sigma}
\epsilon({\bk}) c_{\bk \sigma}$ with $N(\omega) \sim \omega$.

Because of the  vanishing DOS in a d-wave superconducting state
Kondo singlet formation occurs only for a coupling constant
exceeding some critical value~\cite{Kondodwave}. For a
particle-hole symmetric spectrum Kondo singlet is not formed for
arbitrarily large values of $J$. Another situation where Kondo
effect is irrelevant is the case of ferromagnetic coupling $J$.
This allows us, quite generally, to consider a  single spin in a
d-wave superconductor that is not screened and we ignore the Kondo
effect.

In the presence of magnetic field $\mathbf{B}||\hat{\mathbf{z}}$
spin degeneracy is lifted and  components of the spin
$\mathbf{S}||\hat{\mathbf{z}}$ and $\mathbf{S}\perp \mathbf{B}$
will have different propagators. It is obvious that only
transverse components of the spin will contain information about
level splitting  at $\omega_0 = \omega_L$. We have therefore
focused on $S^+, S^-$ components only. The propagator in imaginary
time $\tau$  is $\chi(\tau) = \langle T_{\tau}
S^+(\tau)S^-(0)\rangle$ with Fourier transform and continuing to
real frequency $\chi_{0}(\omega ) = \frac{\langle
S^{z}\rangle}{\omega_0^{2}-(\omega +i\delta)^{2}}$. For free spin
we have $\langle S_z\rangle = \tanh (\omega_0/2T)/2$. For more
general case of  magnetic anisotropy this does not have to be the
case. To be general we will keep $\langle S_z\rangle$.

 We begin with evaluation of the DOS correction due to coupling to
 localized spin.
 Self-energy correction is:
\begin{equation}\label{EQ:Selfenergy1}
  \Sigma(\omega_l) = J^2 T \sum_{\bk,\Omega_n}G(\bk, \omega_l -
  \Omega_n) \chi^{+-}(\Omega_n)\;,
\end{equation}
where $G^0(\bk, \omega_l)  = [i\omega_l - \epsilon(\bk)][
(i\omega_l)^2 - \epsilon^2(\bk) - \Delta^2(\bk)]^{-1} $ is the
particle Green's function in  d-wave superconductor, $G^{-1} =
G^{(0)-1} - \Sigma$,  $F^0(\bk,\omega_l) = [\Delta(\bk)][
(i\omega_l)^2 - \epsilon^2(\bk) - \Delta^2(\bk)]^{-1}$; $\Omega_l
= 2 \pi l T$ is the bosonic Matsubara frequency and $\omega_l =
(2l+1) \pi T; l=0,1,2...$ is the fermionic frequency. Using
spectral representation and analytical continuation onto real axis
$i \omega_n \rightarrow \omega + i\delta $ we find for imaginary
part of self energy $\Sigma(\omega)$ :
\begin{equation}\label{EQ:Selfenergy2}
  \mbox{Im} \Sigma(\omega) = -J^2  \langle S_z\rangle   \mbox{Im}
  G(\omega-\omega_0)[n_F(\omega - \omega_0)- n_B(\omega_0) - 1]\;,
\end{equation}
where $n_F(\omega) = 1/[1+\exp(\beta \omega)], n_B(\omega)=
1/[\exp(\beta \omega)-1]$ are Fermi and Bose distribution
functions. This local self-energy leads to the modifications of
the DOS. In this solution we treat the self-energy effects in G to
all orders, i.e. $G$ in Eq.(\ref{EQ:Selfenergy2}) is  full Green's
function $G^{-1} = G_0^{-1} - \Sigma(\omega)$ and solution for
$\Sigma$ is found self-consistently for  a local vibrational mode.
The modifications of the superconducting order parameter and
bosonic propagator were ignored in this calculation. Results are
presented in Fig.~\ref{FIG:DOScomplete}. To proceed with analytic
treatment, unless stated otherwise, we limit ourselves below to
second order scattering in $\Sigma$. Difference between
self-consistent solution and second order calculation are only
quantitative and small for small coupling. Corrections to the
Green's function $ G({\bf r},{\bf r'},\omega) = G^0({\bf r},{\bf
r'},\omega) + G^0({\bf r},0,\omega) \Sigma(\omega) G^0(0,{\bf r'},
\omega) + F^0({\bf r},0,\omega) \Sigma(\omega) F^{*0}(0,{\bf
r},\omega)$. For simplicity we define $K(T, \omega,\omega_0) =
-[n_F(\omega - \omega_0)- n_B(\omega_0) - 1 \simeq \Theta(\omega -
\omega_0)]$ which becomes a step function at low $T\ll \omega_0$,
the limit we will focus on hereafter. Correction to the local
density of states as a function of position comes from the
correction to the bare Green's function $G^0: \delta N({\bf r},
\omega) = 1/\pi \mbox{Im}[G^0({\bf r},0,\omega) \Sigma(\omega)
G^0(0,{\bf r}, \omega) \pm F^0({\bf r} ,0,\omega)\Sigma(\omega)
F^{*0}(0,{\bf r},\omega) ]$, where keeping it general, the plus
sign corresponds to the coupling to the local vibrational mode and
minus -- to the spin scattering respectively. The strongest effect
will be at the impurity site. For on-site density of states we
have:
\begin{eqnarray}\label{EQ:DOS1}
&&\frac{\delta N({\bf r} = 0, \omega)}{N_0} = \frac{\pi^2}{2}
(JSN_0)^2
\frac{\omega - \omega_0}{\Delta}K(T,\omega,\omega_0)\nonumber\\
&&\times \left(\frac{2\omega}{\Delta}
\ln\left(\frac{\Delta}{\omega}\right)\right)^2, \ \ \omega \ll
\Delta\;,
\end{eqnarray}
\begin{eqnarray}\label{EQ:DOS2}
&&\frac{\delta N({\bf r} = 0, \omega)}{N_0} = 2\pi^2
(JSN_0)^2K(T,\omega,\omega_0) \ln^2\left(\frac{|\omega -
\Delta|}{4\Delta}\right)\nonumber\\
&&\times \ln\left(\frac{4\Delta}{|\omega + \omega_0 -
\Delta|}\right) + (\omega_0 \rightarrow -\omega_0),\ \omega \simeq
|\Delta|\;,
\end{eqnarray}
where we used  for on-site Green's function $G^0(0,0,\omega) =
N_0[ \frac{2\omega}{\Delta}\ln(\frac{4\Delta}{\omega}) + i\pi
\frac{|\omega|}{\Delta}]$, for  $ \omega \ll \Delta$ and  we
retained only dominant real part of $G^0$. In opposite limit
$\omega \simeq \Delta$  we    retained only dominant imaginary
part of $G^0(0,0,\omega) = i\pi N(\omega) = -2iN_0
\ln(\frac{|\omega - \Delta|}{4\Delta})$. At ${\bf r} = 0$ we have
$F^0(0,0,\omega) = 0$. Complete DOS $N(\omega)$ and derivative
$\frac{dN(\omega)}{d\omega}$ are shown on
Fig.~\ref{FIG:DOScomplete}. For arbitrary position $N({\bf
r},\omega)$ we would have to add a Friedel oscillation factor $
\Lambda({\bf r}) = [|G^{0}({\bf r},\omega)|^{2} \pm |F^0({\bf r},
\omega)|^2] \sim \frac{\sin(k_F r)}{(k_Fr_{||})^2 +
(r_{\perp}/\xi)^2}$ that describes the real space dependence of
the Green's function on distance for small $\omega \ll \Delta$.
Here $\mathbf{r}_{\perp}|| \mathbf{k}_{F \perp}$ is the component
of $\bf r = (r_{\perp}, r_{||})$ that is along the Fermi surface
near the nodal point of the gap and $\mathbf{r}_{||}||
\mathbf{k}_{F ||}$ is the component perpendicular  to the Fermi
surface at nodal point. Existence of the nodes in d-wave case
results in the power law decay of $\Lambda({\bf r})$ in all
directions and it has a four fold modulation due to gap anisotropy
(See detailed discussions in PRB '97 reference in \cite{BSR}). The
final result is our Eqs.~(\ref{EQ:1}-\ref{EQ:2}).

\begin{figure}
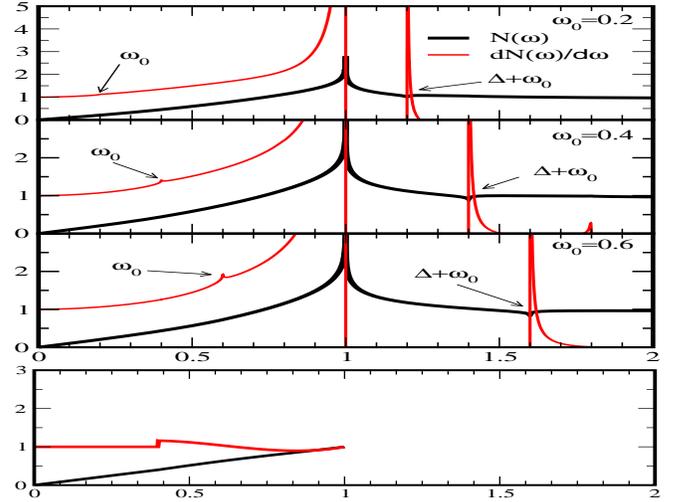

\includegraphics*[width=\columnwidth,height=1.9in]{comparisonnew}
\includegraphics*[width=\columnwidth,height=0.70in]{analytic}
\caption{\label{FIG:DOScomplete} The solution for the DOS (black
line) and its energy derivative (red line) are presented for a
local boson mode scattering in a d-wave superconductor. The normal
self-energy was treated self-consistently as a full solution of
the Eq.~(\ref{EQ:Selfenergy2}), ignoring vertex corrections and
gap modifications. Apart from the feature at $\omega = \omega_0$
we get also strong satellite peaks at $ \Delta + \omega_0$ that
are a consequence of a coherence peak in DOS of a d-wave
superconductor. These satellites are a specific property of a
superconducting state and will not be present in a pseudogap
state. These features are best seen in $\frac{dN}{d\omega}$.
Energy scale is given in units of $\Delta$, the dimensioneless
coupling constant is taken to be 1. For comparison we plot the
results for local mode frequency $\omega_0/\Delta = 0.2, 0.4, 0.6
$ in the first three panels. The lower panel gives the results for
the asymptotic analytic solution, that assumes $\omega_0 \ll
\Delta$ using Eq.(1), for $\omega_0 = 0.4$. The overall features
are similar for both cases, however the analytic solution shows a
somewhat larger feature.}
\end{figure}


It follows immediately that
\begin{eqnarray}\label{EQ:d2Id2V}
&\delta \frac{dI}{dV}/\frac{dI}{dV} \sim \delta N({\bf r} = 0,
V)/N_0 \sim (JSN_0)^2 \frac{V - \omega_0}{\Delta} \Theta(V -
\omega_0)\;,\nonumber\\
&\delta \frac{d^2I}{dV^2} \sim (JSN_0)^2\Theta(V - \omega_0)\;.
\end{eqnarray}
Here we have used the fact that the derivative of
$(\omega-\omega_0)\Theta(\omega - \omega_0)$ with respect to
$\omega$ yields $\Theta(\omega-\omega_0)$. Thus in a d-wave
superconductor and in a metal with vanishing DOS $N(\omega) = N_0
\frac{\omega}{\Delta}$ one should expect a {\em step
discontinuity} in $d^2I/dV^2$ at the energy of a local mode with
the strength $J^2N^2_0$ (see Fig.~\ref{FIG:DOScomplete}). This
result is qualitatively different from the case of conventional
metal. For metal with energy independent DOS we have from
Eq.~(\ref{EQ:DOS1}) for $T\ll\omega_0$
\begin{equation}\label{EQ:DOSMETAL}
\frac{dI}{dV} \sim \delta N({\bf r}=0,V)  \sim J^2 N^3_0 \Theta(V
- \omega_0)\;,
 \end{equation}
 and the second derivative will reveal a delta function $d^2I/dV^2
 \sim J^2 N^3_0 \delta(\omega
- \omega_0) $ The effect in d-wave superconductor is clearly
smaller than correction to DOS in a normal metal with the same
coupling strength.

For completeness we also have calculated the effect of inelastic
scattering in a metal with the more general DOS $N(\omega) = 1/\pi
\mbox{Im} G^0(0,0,\omega) =  (\omega/\Delta)^{\gamma}N_0$ with
power $\gamma > 0$ that is determined by the microscopic
properties of the material. Then, from
Eqs.~(\ref{EQ:Selfenergy2}-\ref{EQ:DOS1}) we have for $\omega \ll
\Delta$:
\begin{eqnarray}\label{EQ:DOSFRAC}
&\delta \frac{dI}{dV}/\frac{dI}{dV} \sim \delta N({\bf r} = 0,
V)/N_0 \sim  (V - \omega_0)^{\gamma} \Theta(V -
\omega_0)\;,\nonumber\\
&\delta \frac{d^2I}{dV^2} \sim (V - \omega_0)^{\gamma -1}\Theta(V
- \omega_0)\;.
\end{eqnarray}
Depending on the value, we get divergent singularity at $\omega_0$
for $\gamma < 1$, or a power law rise for $\gamma \geq 1$. In case
of $\gamma = 1$ we recover the result  for d-wave superconductor
and for a pseudogap normal state.

\begin{figure}[htbp]
\begin{center}
\includegraphics[width = 3.0 in,height=1.5in]{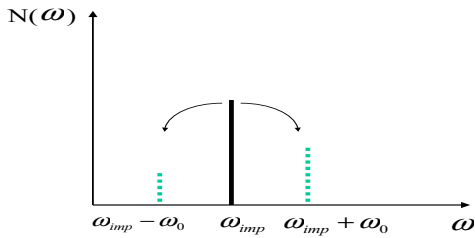}
\caption{Appearance of the satellite peaks for an impurity
resonance $\omega_{imp}$ at $\omega_{imp}\pm \omega_0$ is shown
schematically. The satellites will have different spectral weight.
Imagine we inject into system an electron at energy $\omega_{imp}+
\omega_0 $. To create a peak at $\omega_{imp}$ one need to excite
local mode and the energy of the electron will be equal to the
difference between local state and local mode energies. Similarly,
to obtain the peak at $\omega_{imp}$ from injected electron at
energy $\omega_{imp} - \omega_0$ one needs to add local mode
energy to an electron. For this process to occur the local mode
has to be excited to begin with and hence this process will have
very low weight at low T.  These two processes will also have
different matrix elements. Overall relative weight of the side
peaks is proportional to $J^2N^2_0$ which we assumed to be small.
In case of magnetic scattering, when $\omega_0 = g \mu_B B$ the
splitting will be tunable by the field.} \label{FIG:satellite}
\end{center}
\end{figure}

Quite generally one can express the results in terms of the
spectrum of superconductor.  We can write $ \mbox{Im}
\Sigma(\omega)$ using spectral representation for $G(\mathbf{r},
\omega)$. In superconducting case, using Bogoliubov
$u_{\alpha}(\mathbf{r}), v_{\alpha}(\mathbf{r})$ for eigenstate
$\alpha$, we have $G(\mathbf{r}, \omega) =
\sum_{\alpha}[\frac{|u_{\alpha}(\mathbf{r})|^2}{\omega -
E_{\alpha} + i\delta} + \frac{|v_{\alpha}(\mathbf{r})|^2}{\omega +
E_{\alpha} - i\delta}]$. Taking imaginary part of $G(\mathbf{r},
\omega)$ we arrive  for $T \ll \omega_0$ at:
\begin{eqnarray}
\label{EQ:Selfenergy3} \mbox{Im} \Sigma(\omega) = \frac{\pi J^2}{2
\omega_0} \langle S_z\rangle
[|u_{\alpha}(\mathbf{r}=0)|^2\delta(\omega-\omega_0-E_{\alpha})
\nonumber \\
+|v_{\alpha}(\mathbf{r}=0)|^2\delta(\omega-\omega_0+E_{\alpha})]\;
 \  , \omega >0 .
\end{eqnarray}
At negative $\omega < 0$ one has to replace $\omega_0 \rightarrow
-\omega_0$ in Eq.(\ref{EQ:Selfenergy3}). For example, consider a
magnetic impurity resonance in d-wave superconductor at energy $
\omega_{imp}$, such as a Ni induced
resonance~\cite{BSR,Ninature}. Then only the term with resonance
level $E_{\alpha} = E_{imp}$  will dominate the sum over
eigenstates $\alpha$ in the vicinity of impurity site. Inelastic
scattering off this impurity induced resonance will produce
additional satellite {\em split away from the impurity level by
 $\omega_0$}, see Fig.~\ref{FIG:satellite}.
   Sharp coherence peaks will also produce split satellites.  Again, for a
  local phonon mode one gets a similar
splitting of impurity level with $\omega_0$ now being the phonon
energy.

Our results  suggest the possibility of {\em single  spin
detection} as one monitors the feature in $d^2 I/dV^2$ as a
function of position and external magnetic field. If we take
experimentally seen DOS $N_0 \simeq  1/eV$ with $JN_0 \simeq 0.14,
\Delta = 30 meV$~\cite{Ninature} and assuming the field of $\sim
10 T$ we have $\omega_0 = 1 meV$ (corresponding to the Zeeman
splitting of $\sim 1 meV$ in a magnetic field, we have from
Eqs.~(\ref{EQ:DOS1}-\ref{EQ:d2Id2V})
\begin{equation}\label{EQ:estimate}
 \delta N({\bf r} = 0, \omega)/N_0 \simeq  10^{-2}
\frac{\omega -\omega_0}{\Delta} \Theta(\omega-\omega_0)\;.
\end{equation}
We point out here that result is expressed in terms of the
relative change of DOS of a metal $N_0$. For observation of this
effect one would have to sample DOS in the vicinity of $eV =
\omega_0 \propto B$. Assuming $\omega - \omega_0 = \omega_0$ we
have from Eq.~(\ref{EQ:estimate}) $\delta
\frac{dI}{dV}/\frac{dI}{dV} \sim 10^{-2}$. Expressed as a relative
change of {\em DOS of a superconductor} $N(\omega) = N_0
\omega/\Delta$ effect is: $\delta \frac{dI}{dV}/\frac{dI}{dV} \sim
\delta N({\bf r} = 0, \omega)/N(\omega_0) \sim 10^{-2}
\frac{\omega -\omega_0}{\omega_0} \Theta(\omega-\omega_0)$. It is
of the same order of magnitude as the observed vibrational modes
of localized molecules in inelastic electron tunneling
spectroscopy STM, IETS-STM~\cite{Ho}. The satellites at $ \Delta +
\omega_0$ produce the effect on the scale of unity and clearly
seen even for small coupling.  The important difference is that
for localized spin the kink in DOS is {\em tunable} with magnetic
field and this should make its detection easier.

In conclusion, we propose the extension of the inelastic tunneling
spectroscopy on the strongly correlated electrons states, such as
a d-wave superconductor and pseudogap normal state.  The DOS in
these systems  has a nontrivial energy dependence of general form
$N(\omega) \sim \omega^{\gamma}, \gamma > 0$. This technique could
allow for a Zeeman level spectroscopy of a single magnetic center,
thus, in principle,  allowing a single spin detection. We find the
feature in $dI/dV \sim (\omega - \omega_0)^{\gamma - 1}
\Theta(\omega - \omega_0)$ near the threshold energy $\omega_0$.
We also find strong satellite features near the gap edge due to
coherence peak for a superconducting case. The singularity is a
power law and qualitatively different from the results for a
simple metallic DOS \cite{Ho}. For the relevant values of
parameters for high-Tc the feature is on the order of several
percents and makes the feature observable in these materials.
Similar predictions are also applicable to the local vibrational
modes, where $\omega_0$ becomes a vibrational mode frequency.

{\bf Acknowledgments}: This work was supported by the US
Department of Energy. We are grateful to S. Davis, conversations
with whom initiated this research. We are grateful to A. Chubukov,
E. Hudson, Y. Manassen and D. Scalapino  for useful discussions.
Ar. A. was supported by LDRD 200153, AVB and JXZ were supported by
LDRD X1WX at Los Alamos.


\begin{thebibliography}{99}
\bibitem{Comment1} To minimize the orbital effect of  magnetic field  one can apply
it parallel to the surface of superconductor. The magnetic field
is  penetrating the surface sheath on the scale of penetration
depth so that its effect on the conduction electrons for d-wave SC
is small.
\end{thebibliography}
\end{document}